\documentclass{mecbic}
\providecommand{\anno}{2011}
\providecommand{\firstpage}{7}
\providecommand{\lastpage}{23}

\usepackage{graphicx}
\usepackage{color}
\usepackage{epsfig}
\usepackage{amssymb}
\usepackage{amsmath}
\usepackage{amsthm}
\usepackage[ansinew]{inputenc}
\usepackage{dsfont}
\usepackage{fancybox,tikz}

\newtheorem{proposition}{Proposition}
\newtheorem{definition}{Definition}

\newtheorem{example}{Example}

\usepackage{latexsym}
\usepackage[T1]{fontenc}

\newcommand{\may}{\:\mbox{{\bf may}}\:}
\newcommand{\must}{\:\mbox{{\bf must}}\:}

\newcommand{\bbbn}{{\rm I\!N}}

\newcommand{\RR}{\mathcal{R}}
\newcommand{\II}{\mathcal{I}}

\newcommand{\DD}{\mathcal{D}}

\newcommand{\ltrans}[1]{\xrightarrow{#1}}
\newcommand{\lltrans}[2]{\underset{#1}{\xrightarrow{#2}}}

\newcommand{\qqop}[1]{\mathrel{\makebox[2em]{$#1$}}}

\newcommand{\agr}{\;\;\big|\;\;}

\newcommand{\brane}[3]{[_{#2}\, #1\, ]_{#2}^{#3}}

\newcommand{\flatbrane}[3]{[\![_{#2}\, #1\, ]\!]_{#2}^{#3}}

\newcommand{\genruleD}{\ensuremath{u \rightarrow v_h v_o \{v_{l_i}\}|_D}}

\newcommand{\upout}{\ensuremath{O^\uparrow}}
\newcommand{\downout}{\ensuremath{O^\downarrow}}
\newcommand{\upin}{\ensuremath{\II^\downarrow}}
\newcommand{\downin}{\ensuremath{I^\uparrow}}

\newcommand{\set}{\mathsf{Set}}

\allowdisplaybreaks[1]

\title{A Testing Framework for P Systems}
\author{Roberto Barbuti
\institute{Dipartimento di Informatica, University of Pisa\\
Largo Pontecorvo 3, 56127 Pisa, Italy}
\email{barbuti@di.unipi.it}
\and
Diletta Romana Cacciagrano
\institute{School of Science and Technology, University of Camerino\\
Via Madonna delle Carceri 9, 62032 Camerino (MC), Italy}
\email{diletta.cacciagrano@unicam.it}
\and
Andrea Maggiolo-Schettini
\institute{Dipartimento di Informatica, University of Pisa\\
Largo Pontecorvo 3, 56127 Pisa, Italy}
\email{maggiolo@di.unipi.it}
\and
Paolo Milazzo
\institute{Dipartimento di Informatica, University of Pisa\\
Largo Pontecorvo 3, 56127 Pisa, Italy}
\email{milazzo@di.unipi.it}
\and
Luca Tesei
\institute{School of Science and Technology, University of Camerino\\
Via Madonna delle Carceri 9, 62032 Camerino (MC), Italy}
\email{luca.tesei@unicam.it}}
\def\titlerunning{A Testing Framework for P Systems}
\def\authorrunning{R. Barbuti, D. R. Cacciagrano, A. Maggiolo-Schettini\\\
P. Milazzo and L. Tesei}
\begin{document}
\maketitle
\pagestyle{plain}
\pagenumbering{arabic}
\setcounter{page}{7}

\begin{abstract}
Testing equivalence was originally defined by De Nicola and Hennessy in a
process algebraic setting~(CCS) with the aim of defining an equivalence
relation between processes being less discriminating than bisimulation and with
a natural interpretation in the practice of system development. Finite
characterizations of the defined preorders and relations led to the possibility
of verification by comparing an implementation with a specification in a
setting where systems were seen as black boxes with input and output
capabilities, thus neglecting internal undetectable behaviours.

In this paper, we start defining a porting of the well-established testing
theory into membrane computing, in order to investigate possible benefits
in terms of inherited analysis/verification techniques and interesting
biological applications. P Algebra, a process algebra for describing P Systems,
is used as a natural candidate for the porting since it enjoys the desirable
property of being compositional and comes with other observational equivalences
already defined and studied.

We consider P Systems with multiple membranes, dissolution, promoters and
inhibitors. Notions as {\it  observable} and {\it  test} are conveniently
rephrased in the membrane scenario, where they lack as native notions and have a
not so obvious mean. At the same time, concepts as promoters, inhibitors,
membrane inclusion and dissolution are emphasized and exploited in the attempt
of realizing a testing machinery able to formalize several features, which are
proper of membranes and, as a consequence, worth being highlighted as basic
observables for P Systems. The new testing semantics framework inherits from the
original one the ability to define qualitative system properties.
Moreover, it results to be suitable also to express {\it  quantitative}
aspects, a feature which turns out to be very useful for the biological
domain and, at the same time, puts in evidence an expected high expressive power
of the framework itself.
\end{abstract}

\section{Introduction}

Membrane computing, the research field initiated by Gheorghe P\u aun
\cite{Paun2002,Paun2002a},
aims to define computational models, called {\it  P Systems}, which are inspired
by the
behaviour and structure of the living cell. Since its introduction, the P System
model has been intensively studied and
developed: many variants of membrane systems have been proposed and
regular collective volumes are annually edited.

The most investigated membrane computing topics are related to the computational
power of different variants, their capabilities to
solve hard problems, decidability, complexity aspects and hierarchies of classes
of languages produced by these devices.

In the last years, there have also been significant developments in using the P
systems paradigm for modelling, simulation and
formal verification \cite{ciobanu2006}. Although such topics  have been
exercised
to different classes of P Systems \cite{andrei07}, testing has been quite
neglected in this context.

\subsection{Testing and P Systems}
Testing P Systems has been so far considered by using certain coverage
principles.
More often the rule coverage is utilised, by taking into account different
contexts. Such contexts - typically grammar, automaton and model checking
techniques - are described in depth in \cite{ghe09,ghe10}:
\begin{itemize}
\item[-] ({\it  Grammar-based methods}) In order to test an implementation
developed from a P System specification
in a grammar-based method, a
test set is built, in a black box manner, as a finite set of sequences
containing
references to rules.
Although there are similarities between context-free grammars utilised in
grammar testing and basic P Systems, there are also major
differences that pose new problems in defining testing methods and strategies
to obtain test sets. Some of the difficulties encountered when some grammar-like
testing procedures are introduced, are related to the hierarchical
compartmentalisation of the P System model, parallel behaviour, communication
mechanisms,
the lack of a non-terminal alphabet and the use of multisets of objects
instead of sets of strings.

\item[-] ({\it  Finite-state machine methods}) Finite state machine-based
testing is widely used for software testing. It provides
very efficient and exhaustive testing strategies and well investigated methods
to
generate test sets. In this case it is assumed that a model of the system under
test is provided in the form of a finite state machine. In the P System model
case, such a machine is typically obtained from a partial computation in a P
System.

A different approach uses a special class of state machines,
called {\it  X-machines}. Given that the relationships between various classes
of P Systems and these machines are well studied \cite{aguado02} and the
X-machine-based
testing is well developed, standard techniques for generating test sets based
on X-machines can be adapted to the case of P Systems \cite{ipate09}.

Specific coverage criteria are defined in the case of finite state machine-based
testing. One such criterion,
called {\it  transition coverage}, aims to produce a
test set in such a way that every single transition of the model is covered.

\item[-] ({\it  Model checking-based methods}) The generation of different
test sets, according to certain coverage criteria, can
be done by utilising some specific algorithms or by applying some tools that
indirectly will generate test sets. Such tools, like model checkers, can be used
to
verify some general properties of a model and when these are not fulfilled then
some counter-examples are produced, which act as test sets in certain
circumstances.

In the case of P Systems, an encoding based on a Kripke structure associated
with the system is provided for model checkers like NuSMV \cite{ipate10} or SPIN
\cite{Ipate2011}. This relies on certain operations defined in \cite{dang06} and
encapsulates the main
features of a P System, including maximal parallelism and communication, but
within a
finite space of values associated with the objects present in the system.

The rule coverage principle is expressed by using temporal logics queries
available in such
contexts. By negating specific coverage criteria, counter-examples are
generated.
\end{itemize}

\subsection{Our contribution: a Process Algebra-based testing machinery for P
Systems}

The community of Process Algebra taught us that the usefulness of
formalisms for the description and the analysis of reactive systems is closely
related to the underlying notion of {\it  behavioural equivalence}. Such an
equivalence should formally identify behaviours that are informally
indistinguishable from each other, and at the same time distinguish between
behaviours that are informally different. The authors of \cite{Barbuti2010} go
toward such a direction, proposing some observational equivalences on a suitable
algebraic notation of P Systems.

One way of determining behavioural equivalences is by observing the systems we
are interested in, experimenting on them, and drawing conclusions about the
behaviour of such systems based on what we see, e.g., testing the system. Such
an approach  has been formalized in the Process Algebra setting by a suitable
testing machinery \cite{DeNicola1984}, pivoting on a restricted form of context,
called {\it  observer}.

The way to exercise (evaluate) a process on a given observer is done by letting
the considered process and the given observer to run {\it  in parallel} and by
looking at the computations which the {\it  running test} can perform.

It is worth noting that internal actions of the process under test do not
affect, but the case in which they lead to divergence, the satisfaction of the
test: they are not observable as input or outputs, thus they cannot be perceived
by an external user that is experimenting on the system. This idea is
typical of a testing framework: systems are considered black boxes and only
observables matter in their comparison. This characteristic is imported in the
testing notion introduced in this paper. Internal production and migration of
objects in the P system under test will not be seen by the observer: only the
objects injected into the system by the observer and the objects that are
returned from the system to the observer will matter.

Another typical characteristic of the testing framework, worth underlining, is
that the observer must have the ability to force the system under test to follow
certain paths among all the possible ones. This is in order to investigate, for
instance, what the system can do after a quite specific sequence of inputs, or
after a predefined sequence of inputs/outputs. In order to guarantee this
possibility, and thus giving the equivalence a discriminating power similar to
the one in the original setting, in the testing framework we introduce in this
paper we exploit promoters and inhibitors. Without them the intrinsic nature
of P system behaviour, in particular maximal parallelism, would have prevented
this central feature, thus invalidating the porting.

The characteristics discussed above are central in the idea of testing
equivalence. Bisimulation-based equivalences, even the weak one, are highly
discriminating and do not reflect a practical view of ``testing'' a
system: usually, and \emph{this is always the case for biological
systems}, the whole internal structure/dynamics of the system is not known, but
it is required to check bisimulation. The only way to study such
systems is to interact with them and analyse what can be observed from
experiments, with the means that are available. Along this idea, we start with
this work the definition of a testing framework with the characteristics above,
giving initial theoretical results and some simple examples of tests, without
any particular biological impact. However, we intend as future work to
investigate and exploit the analogy of the defined notion of testing with
biological experiments in order to give more evidence of the biological
relevance of the work. We can devise, for instance, techniques for experimental
planning that could be of great interest for experimental biologists, along the
same line of the techniques proposed in \cite{Ahmad2006}.

In \cite{DeNicola1984}, different equivalence relations (e.g., {\it  may} and
{\it  must} testing equivalences) between systems are defined. Two systems are
considered equivalent if they pass exactly the same set of observers. Such
equivalences are further broken down into preorder relations on systems, i.e.,
relations that are reflexive and transitive (though not necessarily symmetric).
Formally, given a process $P$ and an observer $o$,
\begin{itemize}
\item[-] $P\:\may\: o$ means that there exists a successful computation from
the test $P\: |\:o$
(where $|$ is the parallel operator, and successful means that there is a state
where the special action $\omega$ is enabled);
\item[-] $P\:\must\: o$ means that every maximal computation from
the test $P\: |\:o$ is successful;
\item[-] The preorder $P\leq_{\!\mbox{\it  sat}} Q$ means
that for any observer $o$, $P\:\mbox{\bf sat}\: o$ implies $Q\:\mbox{\bf sat}\:
o$,
where  $\mbox{\bf sat}$  denotes $\mbox{\bf may}$ or  $\mbox{\bf must}$;
\item[-] The equivalence
$P\approx_{\!\mbox{\it  sat}} Q$ means $P\leq_{\!\mbox{\it  sat}} Q$ and
$Q\leq_{\!\mbox{\it  sat}} P$.
\end{itemize}

In \cite{RV07} and in \cite{NC95} a new testing semantics was proposed to
incorporate the fairness notion: the {\it  fair}-testing (aka {{\it 
should}-testing) semantics. In contrast to the classical {\it  must}-testing
(semantics), {\it  fair}-testing abstracts from certain divergences, e.g.,
infinite loops of $\tau$ (invisible) actions. This is achieved by stating that
the observer $o$ is satisfied if success always remains within reach in the
system under observer. In other words, $P \; {\bf fair }\; o$ holds if  in every
maximal computation from $P\: |\: o$  every state can lead to success after
finitely many interactions.

%

On the basis of \emph{P Algebra}, the algebraic notation of P Systems
introduced in \cite{Barbuti2010} (see Section \ref{sec:background}), we adapt
for P Systems the testing machinery defined in \cite{DeNicola1984} (see Section
\ref{sec:framework}).

We  introduce the concept of {\it  context} in case of P Systems expressed as P
Algebra terms. Using this concept, we then define what we consider an {\it 
observer}, which is again a P Algebra term with certain characteristics. This
leads naturally to the definition of computations of a {\it  running test}, i.e.
a tested system running together with an observer. Then, following the classical
definition of \cite{DeNicola1984}, we define the success of a running test in
the two well-known versions of {\it  may} and {\it  must}\footnote{In this paper
we do not consider {\it  fair} testing.}. Finally, the testing preorders are
introduced, together with the induced equivalence relations. More in detail:
\begin{itemize}
\item[-] An observer consists of a membrane structure in which the skin membrane
contains several membranes (with possibly sub-membranes) one of which is a hole,
i.e., a place where another fully defined P
System can be placed and run. The skin membrane of this tested P System
instantiates the hole
membrane and becomes a full component of the running test.

\item[-] P Systems that are
observers are distinguished from P Systems that are normal, testable processes
similarly to the classical testing approach, where a particular action, called
$\omega$, is used to denote the success of a test and, if the running test is
able to
perform this action, then the computation under consideration is a successful
one.  Similarly, in our framework this is easily translated
 introducing a fresh, particular object $\omega$ that, when sent out of the skin
membrane of the running test, denotes the success of the computation.

\item[-] As usual in testing frameworks, we consider only the behaviours of the
running test in which
no output is produced (this corresponds to considering only invisible action
(e.g., $\tau$) computations in a CCS-like Process
Algebra). This is needed to explore all possible behaviours of the tested system
while running together with the observer.
\end{itemize}

An important result consists of the fact that, differently from the
original testing semantics framework, the one proposed here for P Systems
results to be suitable to define both qualitative and (above all)
{\it  quantitative} system properties. This is mainly because the formalism of P
system is expressive enough to express both qualitative and quantitative
aspects. However, these features are crucial for the biological domain. In
Section \ref{sec:esempio}, we put quantitative capabilities in evidence
defining examples of quantitative tests, both concerning time and number of
individuals, of a system modelling a population of individuals that can
reproduce both sexually and asexually. Note that these examples are to
be intended only as explanation of the concepts introduced in the paper and not
as examples of verifications in the model-checking style. The main direction of
continuing this work towards verification is to find finite characterizations of
the resulting testing equivalence, possibly with respect to suitable classes of
observers, and thus using them to compare the expected behaviour of a
finite state system with its actual behaviour.

\section{Background}\label{sec:background}

We first briefly recall the definition of P Systems \cite{Paun2002,Paun2002a}.
Then, we give the definition of the syntax and the semantics of the P Algebra as
it was presented in \cite{Barbuti2010}, where a class of P Systems including
rule promoters and inhibitors \cite{Bottoni2002} was considered. The original
formulation of the P Algebra, without promoters and inhibitors, can be found in
\cite{Barbuti2008,Barbuti2008a} that we refer as a more detailed presentation
of the semantics.

\subsection{P Systems with Promoters and Inhibitors}

A P System consists of a \emph{hierarchy of membranes}, each of them containing
a multiset of \emph{objects}, representing molecules, a set of \emph{evolution
rules}, representing chemical reactions, and possibly other membranes. Each
evolution rule consists of two multisets of objects, describing the reactants
and the products of the chemical reaction. A rule in a membrane can be applied
only to objects in the same membrane. Some objects produced by the rule remain
in the same membrane, others are sent \emph{out} of the membrane, others are
sent \emph{into} the inner membranes (assumed to exist) which are identified by
their labels.

In the original definition of P Systems, rules are applied with
\emph{maximal parallelism}, namely it cannot happen that a rule is not applied
when the objects needed for its triggering are available. Here, we assume that
at each step at least one evolution rule in the whole system is applied, and
also that more than one rule and several occurrences of the same rule can be
applied at the same step (to different objects). In other words, we assume that
at each step a multiset of evolution rule instances is non-deterministically
chosen and applied in each membrane, such that in the whole system at least one
rule is applied. This is a general form of parallelism that is better suited
than the maximal one to describe events in biological systems.

In P Systems with  promoters and inhibitors an evolution rule in a membrane may
have some
\emph{promoters} and some \emph{inhibitors}. Promoters are objects that are
required to be present and inhibitors are objects that are required to be absent
in the membrane $m$ in order to enable the application of the rule. Promoters
will be denoted simply as objects, namely $a,b,c,\ldots$, while inhibitors will
be denoted as objects preceded by a negation symbol, namely $\neg a, \neg b,
\neg c, \ldots$.

We denote with $\DD_V$ the set of all possible promoters and
inhibitors symbols that can be obtained from an alphabet $V$, namely $\DD_V = V
\cup \neg V$. Given a set of promoter and inhibitor symbols $D$, we denote with~$D^+$ and $D^-$ the sets of objects containing all the objects occurring in $D$
as promoters and all the objects occurring in $D$ as inhibitors, respectively.
We remark that $D^+$ and $D^-$ are sets of objects, hence elements on $D^-$ will
not be preceded by $\neg$. Moreover, with $\neg D$ we denote the set obtained by
transforming each promoter in $D$ into an inhibitor and vice versa. As an
example, if $D=\{a,\neg b, \neg c, d\}$ we have $D^+ = \{a,d\}$, $D^-=\{b,c\}$
and $\neg D = \{\neg a, b, c,\neg d\}$.

We assume that all evolution rules have the following form, where
$u,v_h,v_o,v_1,\ldots,v_n$ are multisets of objects, $\{l_1,\dots,l_n\}$ is a
set of membrane labels in $\bbbn$, and $D$ is a set of promoters and inhibitors:
\begin{equation*}
u \rightarrow (v_h,here) (v_o,out) (v_1,in_{l_1})\dots(v_n,in_{l_n})|_D\, .
\end{equation*}

A rule can be applied only if requirements expressed by $D$ are
satisfied. When a rule is applied, the multiset of objects $u$ is replaced by
$v_h$, multiset $v_o$ is sent to the parent membrane, and each $v_i$ is sent
to inner membrane $l_i$.
Promoters are not consumed by the application of the corresponding evolution
rule and a single occurrence of a promoter may enable the application
of more than one rule in each evolution step. Similarly, a single occurrence of
an inhibitor forbids the application of all the evolution rules in which it
appears. We assume that the set of promoters and inhibitors $D$ of an evolution
rule does not contain the same object both as a promoter and as an inhibitor,
namely $D^+ \cap D^- = \varnothing$, and that consumed objects $u$ are not
mentioned among inhibitors, namely $u \cap D^- = \varnothing$.

\begin{definition}A \emph{P System} $\Pi$ is a tuple $(V , \mu , w_1 , \ldots ,
w_n , R_1 , \ldots , R_n)$ where:
\begin{itemize}
\item $V$ is an \emph{alphabet} whose elements are called \emph{objects};
\item $\mu \subset \bbbn \times \bbbn$ is a \emph{membrane structure}, such
that $(l_1,l_2) \in \mu$ denotes that the membrane labelled by $l_2$ is
contained
in the membrane labelled by $l_1$;
\item $w_j$ with $1 \leq j \leq n$ are multisets of objects in $V$ associated
with the membranes $1,\ldots,n$ of $\mu$;
\item $R_j$ with $1 \leq j \leq n$ are finite sets of \emph{evolution rules}
associated with the membranes $1,\ldots,n$ of $\mu$.
\end{itemize}
\end{definition}
%
%
%
%
\subsection{The P Algebra: Syntax and Semantics}

In this section we recall the \emph{P Algebra}, the algebraic notation of P
Systems we have introduced in \cite{Barbuti2008}, with slight modifications
introduced in \cite{Barbuti2010}. We
assume $V$ to be an alphabet of objects and we adopt the usual string
notation to represent multisets of objects in $V$. For instance, to
represent $\{a,a,b,b,c\}$ we may write either $aabbc$, or $a^2b^2c$, or
$(ab)^2c$. We denote with $\set(u)$ the support of multiset
$u$, namely the set of all the objects occurring in $u$.
We denote multiset (and set) union as string concatenation, hence
we write $u_1u_2$ for $u_1 \cup u_2$. Moreover, we shall write $u(a)$ for the
number of occurrences of $a$ in multiset $u$. For the sake of legibility, we
shall write $\genruleD$ for the generic evolution rule $u
\rightarrow (v_h,here)(v_o,out)(v_1,in_{l_1})\ldots(v_n,in_{l_n})|_D$.

The abstract syntax of the P Algebra is defined as follows.

\begin{definition}[P Algebra] The abstract syntax of \emph{membrane contents}
$c$, \emph{membranes} $m$, and \emph{membrane systems} $ms$ is given by the
following grammar, where $l$ ranges over  $\bbbn$ and $a$ over $V$:
\begin{align*}
c \qqop{::=} & ( \varnothing , \varnothing ) \agr (\genruleD , \varnothing)
\agr (\varnothing , a) \agr c \cup c\\
m \qqop{::=} & \brane{c}{l}{} \\
ms \qqop{::=}& ms \mid ms \agr \mu(m,ms) \agr F(m)
\end{align*}
\end{definition}

A membrane content $c$ represents a pair $(\RR,u)$, where $\RR$ is a set of
evolution rules and $u$ is a multiset of objects.
A membrane content is obtained through the union operation $\_ \cup \_$ from
constants representing single evolution rules and single objects, and can be
plugged into a membrane with label $l$ by means of
the operation $\brane{\_}{l}{}$ of membranes
$m$. Hence, given a membrane content $c$ representing the pair
$(\RR,u)$ and $l \in \bbbn$, $\brane{c}{l}{}$ represents the
membrane having $l$ as label, $\RR$ as evolution rules and $u$ as objects.

Membrane systems $ms$ have the following meaning: $ms_1 \mid ms_2$ represents
the juxtaposition of $ms_1$ and $ms_2$, $\mu(m,ms)$ represents the hierarchical
composition of $m$ and $ms$, namely the containment of $ms$ in $m$, and $F(m)$
represents a \emph{flat membrane}, namely it states that $m$ does not contain
any child membrane.
Juxtaposition is used to group sibling membranes, namely membranes all having
the same parent in a membrane structure. This operation allows hierarchical
composition $\mu$ to be defined as a binary operator on a single membrane (the
parent) and a juxtaposition of membranes
(all the children).

Note that every P System has a corresponding membrane system in the P Algebra,
and that there exist membrane systems which do not correspond to any P System.

In what follows we will often write $\flatbrane{c}{l}{}$ for
$F(\brane{c}{l}{})$. We shall also often write
$(\RR,u)$ where $\RR=\{r_1,\ldots,r_n\}$ is a set of rules and $u = o_1\ldots
o_m$ a multiset of objects rather
than $(r_1,\varnothing)\cup \ldots \cup (r_n,\varnothing)\cup
(\varnothing,o_1)\cup \ldots \cup (\varnothing,o_m)$. Moreover, we shall often
omit parentheses around membrane contents.

The semantics of the P Algebra is given as a labelled transition system
(LTS). The labels of the LTS can be of the following forms:
\begin{itemize}
\item $(u,v,v',D,I,\upout,\downout)$,
describing a computation step performed by a membrane content $c$, where:
\begin{itemize}
\item $u$ is the multiset of objects consumed by the application of
evolution rules in $c$, as it results from the composition, by means of
$\_ \cup \_$, of the
constants representing these evolution rules.
\item $v$ is the multiset of objects in $c$ offered for the application of
the evolution rules, as it results from the composition,
by means of $\_ \cup \_$,
of the constants representing these objects.
When operation $\brane{\_}{l}{}$ is applied to $c$, it is
required that $v$ and $u$ coincide.
\item $v'$ is the multiset of objects in $c$ that are not used to apply any
evolution rule and, therefore, are not consumed, as it results from the
composition, by means of $\_ \cup \_$, of the constants representing these
objects.
\item $D$ is a set of promoters and inhibitors required to be present and
absent, respectively, by the application of evolution rules in $c$. More
precisely, $D^-$ contains all the inhibitors of the applied evolution rules in
$c$, whereas $D^+$ is a subset of the promoters of those rules. Such a subset
contains only those objects that are not present in the multiset of objects of
$c$.
\item $I$ is the multiset of objects received as
input from the parent membrane and from the child membranes.
\item $\upout$ is the multiset of objects sent as an output to the parent
membrane.
\item $\downout$ is a set of pairs $(l_i,v_{l_i})$ describing the multiset of
objects sent as an output to each child membrane $l_i$.
\end{itemize}
\item $(\upin,\downin,\upout,\downout,app)$, describing a computation step
performed by a membrane $m$, where: $\upin$ is a set containing
only the pair $(l,I)$ where $l$ is the label of $m$ and $I$ is the multiset of
objects received by $m$ as input from the parent membrane, $\downin$ is the
multisets of
objects received from the child membranes of $m$, and $\upout$
and $\downout$ are as in the previous case. Finally, $app \in \{0,1\}$ is
equal to $0$ if no rule has been applied in $m$ in the described computation
step, and it is equal to $1$ otherwise.
\item $(\upin,\upout,app)$, describing a computation step performed by a
membrane
system $ms$, where $\upin$, $\upout$ and $app$ are as in the previous cases.
\end{itemize}


For the sake of legibility, in transitions with labels of the first form we
shall write the first four elements of the label under the arrow denoting the
transition and the other elements over the arrow. Now, LTS transitions are
defined through SOS rules \cite{Plotkin2004}. We give here a very short
explanation of such rules. Please, refer to \cite{Barbuti2008} for more details.

\begin{figure}[t]
\framebox{
\begin{minipage}{0.95\textwidth}
\small
\begin{gather*}
\frac{I \in V^* \qquad n \in \bbbn}{
( \genruleD , \varnothing )
\lltrans{u^n,\varnothing,\varnothing,D}{I,v_o^n,\{(l_i,v_{l_i}
^n)\} }
( \genruleD , I v_h^n )
}
\qquad (mc1_n)
\\[0.4\baselineskip]
\frac{I \in V^* \qquad D' \subseteq \neg D \qquad D' \neq \varnothing}{
( \genruleD , \varnothing )
\lltrans{\varnothing,\varnothing,\varnothing,D'}{
\quad I ,\varnothing,\varnothing \quad}
( \genruleD , I )
}
\qquad (mc2)
\\[0.4\baselineskip]
%
%
%
\frac{I \in V^*}{
( \varnothing , a )
\lltrans{\varnothing,a,\varnothing,\varnothing}{\quad I,
\varnothing,\varnothing\quad}
( \varnothing , I )
}
\qquad (mc3)
\qquad \qquad
%
\frac{I \in V^*}{
( \varnothing , a )
\lltrans{\varnothing,\varnothing,a,\varnothing}{\quad I,
\varnothing ,\varnothing\quad}
( \varnothing , I a )
}
\qquad (mc4)
%
\\[0.2\baselineskip]
\frac{I \in V^*}{
( \varnothing , \varnothing )
\lltrans{\varnothing,\varnothing,\varnothing,\varnothing}{
\quad I,\varnothing, \varnothing\quad}
( \varnothing , I )
}
\qquad (mc5)
\\[0.4\baselineskip]
\frac{
\begin{matrix}
x_1 \lltrans{u_1,v_1,v_1',D_1}{I_1,\upout_1,\downout_1} y_1 \qquad
x_2 \lltrans{u_2,v_2,v_2',D_2}{I_2,\upout_2,\downout_2} y_2 \qquad
\\
(D_1^- \cup D_2^-) \cap \set(v_1v_1'v_2v_2') = \varnothing \qquad
D_1 \cap \neg D_2 = \varnothing \qquad
D = ( D_1 D_2 ) \setminus \set(v_1v_1'v_2v_2')
\end{matrix}
}{
x_1 \cup x_2 \lltrans{u_1u_2, v_1 v_2, v_1' v_2', D}{
I_1  I_2,
\upout_1 \upout_2, \downout_1 \cup_{\bbbn} \downout_2}
y_1 \cup y_2
}
\qquad (u1)
\end{gather*}
\normalsize
\end{minipage}
}
\caption{Transition rules for membrane contents and unions
of membrane contents.}\label{fig:rules-mc}
\end{figure}

We start by giving in Fig.~\ref{fig:rules-mc} the transition rules for
membrane contents.
Rule $(mc1_n)$ describes $n$ simultaneous applications of an evolution rule for
any $n \in \bbbn$.
Rule $(mc2)$ describes the case in which an evolution rule is not applied
because a subset $D'$ of the promoters and inhibitors in $D$ it requires to be
present and absent, respectively, are assumed not to satisfy the requirements.
Rules $(mc3),(mc4)$ and $(mc5)$ describe the transitions performed by
membrane contents consisting of a single object and the transitions performed
by an empty membrane content.

\begin{figure}[t]
\framebox{
\begin{minipage}{0.95\textwidth}
\small
\begin{gather*}
\frac{
x \lltrans{u,u,v',D}{I,\upout,\downout} y
\qquad
app=
\begin{cases}
0 & \mbox{if } u = \varnothing\\
1 & \mbox{otherwise}
\end{cases}
\qquad
\begin{matrix}
D^+ = \varnothing
\end{matrix}
}{
\brane{x}{l}{}
\ltrans{
\varnothing , I , \upout , \downout , app} \brane{y}{l}{}
}
\qquad (m1)
\\[0.4\baselineskip]
 \frac{
x \lltrans{u,u,v',D}{I_1I_2,\upout,\downout} y
\qquad
app=
\begin{cases}
0 & \mbox{if } u = \varnothing\\
1 & \mbox{otherwise}
\end{cases}
\qquad
\begin{matrix}
D^+ = \varnothing
\qquad I_1 \neq \varnothing
\end{matrix}
}{
\brane{x}{l}{}
\ltrans{\{(l,I_1)\},I_2,\upout,
\downout,app} \brane{y}{l}{}
}
\quad (m2)
\\[0.4\baselineskip]
\frac{
x \ltrans{\upin,\varnothing,\upout,\varnothing,app} y
}{
F(x) \ltrans{\upin,\upout,app} F(y)
}
\quad (fm1)
%
\qquad \qquad
\frac{
x_1 \ltrans{\II_1,\upout_1,app_1} y_1 \qquad
x_2 \ltrans{\II_2,\upout_2,app_2} y_2
}{
x_1|x_2 \ltrans{\II_1 \II_2,\upout_1 \upout_2, max(app_1,app_2)}
y_1|y_2
}
\quad (jux1)
\\[0.4\baselineskip]
\frac{
x_1 \ltrans{\upin_1,\downin_1,\upout_1,\downout_1,app_1} y_1 \qquad
x_2 \ltrans{\II_2,\upout_2,app_2} y_2 \qquad
\downout_1 \bumpeq \II_2 \qquad \upout_2 = \downin_1
}{
\mu(x_1,x_2) \ltrans{\upin_1,\upout_1,max(app_1,app_2)}
\mu(y_1,y_2)
}
\quad (h1)
\end{gather*}
\normalsize
\end{minipage}
}
\caption{Rules for individual membranes and hierarchical
composition of membranes}\label{fig:rules-memb}
\end{figure}

Rule $(u1)$ describes the behaviour of a union of membrane contents. In this
transition rule we use some auxiliary notations.
Given two sets
$\downout_1$ and $\downout_2$ representing two outputs to inner membranes, we
write $\downout_1 \cup_{\bbbn} \downout_2$ to denote the set
$\{(l,uv) \,| \, (l,u) \in \downout_1 \wedge (l,v) \in \downout_2\}
\cup
\{(l,u) \,| \, (l,u) \in \downout_1 \wedge \nexists v. (l,v) \in
\downout_2\}
\cup
\{(l,v) \,| \, (l,v) \in \downout_2 \wedge \nexists u. (l,u) \in
\downout_1\}$.



In Fig.~\ref{fig:rules-memb} we give transition rules for individual membranes,
juxtaposition and hierarchical composition. Note that from the transition label
of the membrane content we have no information about the objects that have been
produced by the applied evolution rules.
Rules $(m1)$ and $(m2)$ describe the transitions performed by a membrane
with label $l$. In particular, $(m1)$ describes the case in
which no objects are received as an input from the external membrane, while
$(m2)$ describes the case in which a multiset of objects $I_1
\neq \varnothing$ is received.
In these rules $app$ is set to zero if no evolution rule is applied
($u=\varnothing$), and it is set to one if at least one rule is applied
($u\neq\varnothing$).
Rule $(fm1)$ allows us to infer the behaviour of a flat membrane
$\flatbrane{c}{l}{} = F(\brane{c}{l}{})$ from the behaviour of
membrane $\brane{c}{l}{}$.
Rule $(jux1)$ allows us to infer the behaviour of a juxtaposition of
two membrane structures from the behaviours of the two structures.
Finally, rule $(h1)$ describes the behaviour of a hierarchical composition of
membranes. In this rule we assume $\bumpeq$ to be an equivalence relation on
sets of pairs $(l,u)$ with $l \in \bbbn$ and $u \in V^*$, such that, given two
such sets $\II_1$ and $\II_2$, then $\II_1 \bumpeq \II_2$ holds if and only if
$(\II_1 \setminus \{(l,\varnothing) \mid l \in \bbbn\}) = (\II_2 \setminus
\{(l,\varnothing) \mid l \in \bbbn\})$.
In the last two rules $app$ is set to one if at least one between $app_1$ and
$app_2$
is equal to one, namely $app = max(app_1,app_2)$. This means that at least one
rule has been applied in the whole composition.


We conclude by defining a {\it  system trace} as a sequence of internal
information given by an execution of a P Algebra term. We assume that the
system can send objects out of the outmost membrane, but cannot receive objects
from outside. This requirement corresponds to the fact that in a P System
objects cannot be received by the outmost membrane from the external
environment.
Note that executions containing steps in which no rule is applied, namely those
with 0 as last element of the label, are not considered.

\begin{definition}[Trace] A {\it  trace} of a membrane system $ms$ with alphabet
$V$ is a (possibly infinite) sequence $w$ of outputs such that,
for any $\upout_i$ and $ms_i$ with $i \in \bbbn^+$
\begin{itemize}

\item $w=\upout_1 \upout_2 \cdots
\upout_n$ and $ms \ltrans{\varnothing,\upout_1,1} ms_1
\ltrans{\varnothing,\upout_2,1} \ldots
\ltrans{\varnothing,\upout_{n},1} ms_n
\hspace{0.5cm}\not\hspace{-0.5cm}\ltrans{\varnothing,\upout,1}$ \\
\ \\
or
\item $w=\upout_1 \upout_2 \cdots
\upout_n \cdots$ and
$ms \ltrans{\varnothing,\upout_1,1} ms_1
\ltrans{\varnothing,\upout_{2},1} ms_2
\ltrans{\varnothing,\upout_{3},1} ....$ .
\end{itemize}

\noindent We denote with ${\cal T}$ the set of all traces.
\end{definition}

\section{Testing framework}\label{sec:framework}

We first introduce the concept of {\it  context} in case of P Systems
expressed as P Algebra terms. Using this concept, we then define what we
consider an {\it  observer}, which is again a P Algebra term with certain
characteristics. This leads naturally to the definition of computations of a
running test, i.e. a tested system running together with an observer. Then,
following
the classical definition of \cite{DeNicola1984}, we define the success of a test
in the two well-known versions of \emph{may} and \emph{must}. Finally, the
testing preorder is introduced, together with the induced equivalence relation.

\subsection{Contexts and test satisfaction}

At a first look, a natural candidate for context of a P System is another P
System, which we call \emph{observer}, consisting of a membrane structure in
which the skin membrane contains several membranes (with possibly sub-membranes)
one of which is a ``hole'', i.e., a place where another fully defined P System
can be placed and run. The skin membrane of this \emph{tested} P System
instantiates\footnote{This instantiation process may require some trivial
modifications of the tested P System, such as $\alpha$-conversion of the numbers
assigned to the skin and inner membranes.} the ``hole'' membrane and becomes a
full component of the \emph{running test}.

However, in \cite{Barbuti2008a}, a result regarding flattening P
Systems in P Algebra is presented. A similar result can be found in
\cite{Bianco2006}. The flattening process of \cite{Barbuti2008a} reduces any P
System, specified in P Algebra with promoters and inhibitors, into a flat one
(i.e., with no internal membranes) that is bisimilar to the original one. The
notion of bisimulation is the one, based on computation steps, defined in
\cite{Barbuti2008}. These results suggested us to simplify, without loss of
generality, the notion of context we are defining. For this reason, instead of
considering an observer of the form $\mu(m, ms_1 \mid ms_2 \mid \cdots \mid
ms_n \mid \Box)$, we will always consider the equivalent observer
$\mu(m',\Box)$ where $m'$ results from the flattening process of $\mu(m, ms_1
\mid ms_2 \mid \cdots \mid ms_n)$.

Another ingredient of the testing framework is needed to distinguish formally P
Systems that are \emph{observers} from P Systems that are normal,
\emph{testable} processes. In classical testing a particular action, usually
called $\omega$, is used to denote the ``success'' of a test, that is to say,
if the \emph{running test} is able to perform this action then the computation
under consideration is a successful one. This easily translates in our
framework: we introduction a fresh, particular object, $\omega \not \in
V$ that, when sent out of the skin membrane of the running test, denotes the
``success'' of the computation.

\begin{definition}[Observer]
Let $V$ be the alphabet of objects and let $\omega \not \in V$ be a particular
object. An \emph{observer system}, or simply a \emph{test}, is a P Algebra term
of the form $\mu(m,\Box)$ where $m$ is the skin membrane $\brane{c}{1}{}$,
$\Box$ is an unspecified membrane $ms$ numbered 2 and each rule of $m$ is of the
form $u \rightarrow v_h \omega_o \{v_{l_2}\}|_D$. In other words, $m$
communicates with its only child membrane 2 and can send out only $\omega$
objects.
\end{definition}

\begin{definition}[Running test]
Let $V$ be the alphabet of objects and let $\omega \not \in V$ be a particular
object. Let $\mu(m,\Box)$ be an observer system and let $ms$ be term in P
Algebra denoting a closed membrane (i.e., of the form $F(-)$ or
$\mu(-,-)$). The \emph{running test} is the P Algebra term $\mu(m,ms')$ where
$ms'$ is the term $ms$ in which the former skin membrane (numbered 1) is
re-labelled in 2 and the other internal membranes numbers are $\alpha$-converted
in order not to collide with 1 and 2 (the re-labelling of the membrane
names of course implies also applying the substitutions to all the references to
the membrane names in the term.)
\end{definition}

\begin{definition}[Computations]
Let $\mu(m,ms)$ be a running test. A computation $c$ of $\mu(m,ms)$ is any
sequence of the form:
\begin{itemize}
  \item $c$ is finite: $\mu(m,ms) = ms_0 \ltrans{\varnothing,\varnothing,1} ms_1
\ltrans{\varnothing,\varnothing,1} \ldots
\ltrans{\varnothing,\varnothing,1} ms_n
\hspace{0.5cm}\not\hspace{-0.5cm}\ltrans{\varnothing,\varnothing,1}$\\

or\\
\item $c$ is infinite: $\mu(m,ms) = ms_0 \ltrans{\varnothing,\varnothing,1} ms_1
\ltrans{\varnothing,\varnothing,1} ms_2
\ltrans{\varnothing,\varnothing,1} ....$\\
\end{itemize}
A computation is called \emph{successful} if there are $k \in \bbbn$ and $n
\in \bbbn^+$ such that $ms_k \ltrans{\varnothing,\{\omega^n\},1}
ms_{k+1}$, i.e., at least a success symbol can be sent out of the skin membrane
along the run. A membrane like $ms_k$, from which a transition can be taken
that sends out at least one $\omega$ object is called \emph{success}
membrane. We may also write $ms_k \ltrans{\omega}$ to
indicate that $ms_k$ is a success membrane.
\end{definition}
Note that, as usual in testing frameworks, we consider only the behaviours of
the running test in which no output is produced (this corresponds to
considering only
$\tau$ computations in a CCS-like Process Algebra). This is needed to explore
all possible behaviours of the tested system while running together with the
observer. Note that an unsuccessful computation, i.e.\ one with no
success state along it, can be either finite or infinite. In testing
theories \emph{divergence} must have a delicate treatment because it can lead to
different testing preorders \cite{DeNicola1984,Boreale1999}. For simplicity we
port here the original formulation of \cite{DeNicola1984}. Thus, we define a
unary predicate $ms\uparrow$ meaning that, with respect to success, the
``state'' $ms$ along a computation is ``underdefined'' because it leads to
a divergent computation.

\begin{definition}[Divergence]
  Let $c$ be a computation of a running test $\mu(m,ms)$. $c$ is
\emph{divergent}, which we denote by $c\Uparrow$ iff:
\begin{itemize}
  \item $c$ is unsuccessful, or
\item $c$ contains a membrane $ms$ such that $ms\uparrow$ and
there is not a success membrane $ms'$ preceding $ms$ in $c$.
\end{itemize}
\end{definition}

We now have all the ingredients to define the success of a test. Following the
classical approach, such a definition comes into version: a \emph{may}
satisfaction, weaker, and a \emph{must} satisfaction, stronger.

\begin{definition}[Satisfaction of a test]
  Let $\mu(m,\Box)$ be a test and $ms$ be a closed membrane. Then we say:
\begin{itemize}
  \item $ms$ \may $\mu(m,\Box)$ iff there exists a
computation $c$ of $\mu(m,ms)$ and $k \in \bbbn$ such that $$c: \mu(m,ms) = ms_0
\ltrans{\varnothing,\varnothing,1} ms_1
\ltrans{\varnothing,\varnothing,1} \cdots
\ltrans{\varnothing,\varnothing,1} ms_k$$ and $ms_k \ltrans{\omega}.$
\item $ms$ \must $\mu(m,\Box)$ iff for each computation
$c$ of $\mu(m,ms)$, $$c:\mu(m,ms) = ms_0 \ltrans{\varnothing,\varnothing,1} ms_1
\ltrans{\varnothing,\varnothing,1} ms_2
\ltrans{\varnothing,\varnothing,1} \cdots $$ the following conditions hold:
  \begin{itemize}
    \item[(i)] there is $n \in \bbbn$ such that $ms_n \ltrans{\omega}$
    \item[(ii)] if there is $k\in \bbbn$ such that $ms_k\uparrow$, then
there exists $k' \leq k$ such that $ms_{k'} \ltrans{\omega}.$
  \end{itemize}
\end{itemize}

\end{definition}

Note that, in case of {\it  must} satisfaction, computations can be infinite, but
it
is required that a success state is present just at the beginning of
divergence, or before.

\subsection{Testing preorders and testing equivalences}

Using the definitions of satisfaction of a test, we naturally derive preorders
between membrane systems.

\begin{definition}
  Let $ms_1$ and $ms_2$ two closed membrane systems. We define two
relations $\sqsubseteq_T^m$ and $\sqsubseteq_T$ between closed membrane systems
as follows:
\begin{itemize}
  \item $ms_1 \sqsubseteq_T^m ms_2$ \textbf{iff} for each observer
$\mu(m,\Box$), $ms_1$ \may $\mu(m,\Box)$ $\Rightarrow$ $ms_2$
$\may$ $\mu(m,\Box)$
\item $ms_1 \sqsubseteq_T ms_2$ \textbf{iff} for each observer $\mu(m,\Box$),
$ms_1$ $\must$ $\mu(m,\Box)$ $\Rightarrow$ $ms_2$ \must
$\mu(m,\Box)$
\end{itemize}
\end{definition}

\begin{proposition}
  The relations $\sqsubseteq_T^m$ and $\sqsubseteq_T$ are preorders.
\end{proposition}

\begin{definition}[Testing equivalence]
We say that $ms_1$ is {\bf may} testing equivalent to $ms_2$, and write $ms_1
\approx_T^m ms_2$, iff $ms_1 \sqsubseteq_T^m ms_2$ and $ms_2
\sqsubseteq_T^m ms_1$.

\noindent
Analogously, we say that  $ms_1$ is {\bf must} testing equivalent to $ms_2$,
and write $ms_1
\approx_T ms_2$, iff $ms_1 \sqsubseteq_T ms_2$ and $ms_2 \sqsubseteq_T
ms_1$.
\end{definition}
Since the defined relations are kernels of preorders, it is easy to
conclude that they are equivalence relations.

In \cite{Barbuti2008} some equivalence relations are defined between the terms
of P Algebra. Among them, we consider bisimulation, denoted by $\approx$, and
trace equivalence, denoted by $\approx_{Tr}$. We show that the relationships
between these two equivalences with the {\it  must} testing equivalence are the
same
that hold when classical process calculi, as CCS, are considered
\cite{DeNicola1984}.

\begin{proposition} Let $\approx, \approx_T$ and $\approx_{Tr}$ as above. Then,
  $\approx \;\; \subsetneq \;\; \approx_T \;\; \subsetneq \;\; \approx_{Tr}$.
\end{proposition}
\begin{proof}
  The two set inclusions can be proved rephrasing the argument used for CCS
\cite{DeNicola1984}.  Regarding the strictness of the
inclusions, Figure~\ref{fig:tracetest} shows two systems that are trace
equivalent, but can be distinguished by the shown observer. Not that both
systems
perform the set of traces
$\{\varnothing(\alpha)(\beta),\varnothing(\alpha)(\gamma) \}$. However,
considering the {\it  must} testing, the observer shown in
Figure~\ref{fig:tracetest}$(b)$
is not satisfied by the system $(a)$ because in one computation, after $\alpha$,
only $\gamma$ can be produced, thus that computation is unsuccessful.
Conversely, for the other system $(c)$ all computations are successful. Note
that inhibitors and promoters are used in this settings to select only specific
paths in the system. In the classical setting of testing for synchronous
calculi this role is played by the parallel operator together with the
restriction on observing only $\tau$-computations.

Figure~\ref{fig:testbisim} shows two systems that are test equivalent, but
not bisimilar. Below each system the graph of possible transitions\footnote{For
the sake of legibility, we omit the third field in the transition labels.} is
depicted,
in order to show that bisimulation does not hold.
\end{proof}


\begin{figure}
\begin{center}
  \includegraphics{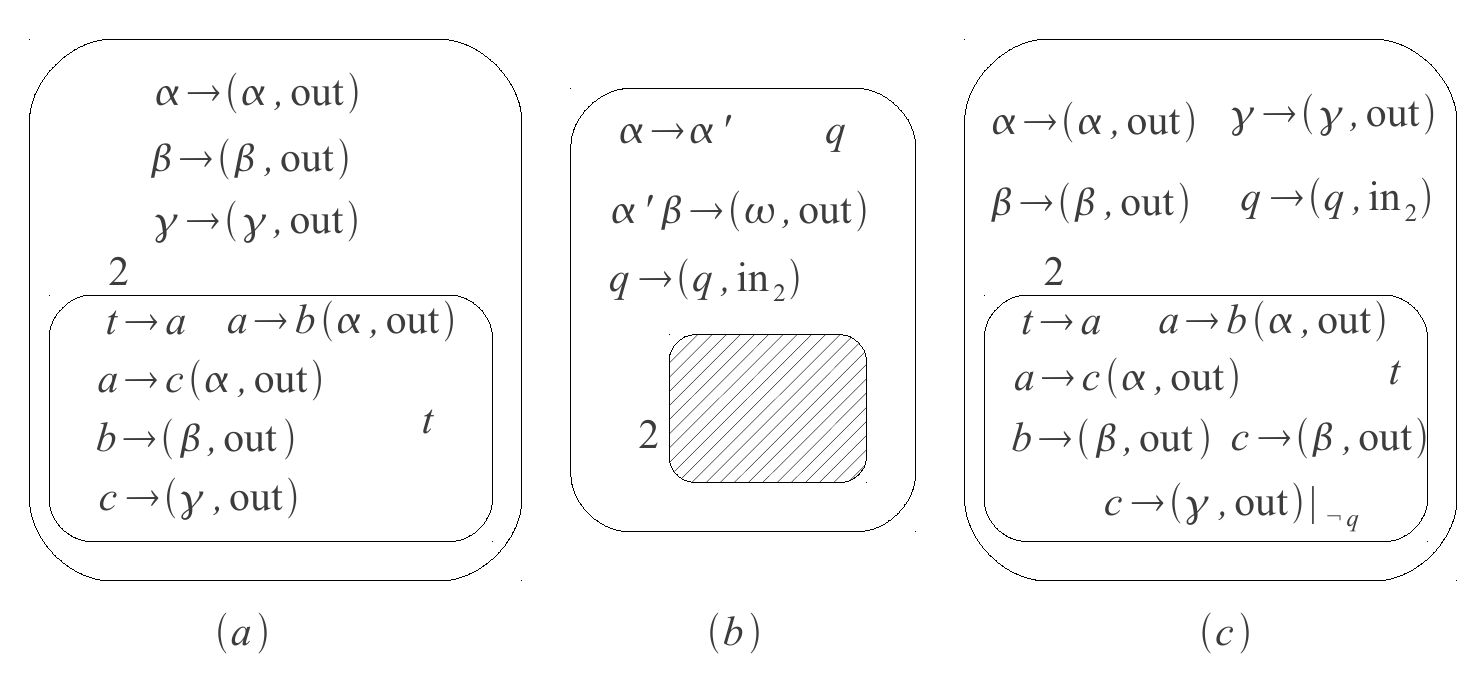}
\end{center}
\caption{Two membrane systems, $(a)$ and $(c)$, that are trace equivalent, but
not test equivalent, together with the observer $(b)$ that distinguishes them.}
\label{fig:tracetest}
\end{figure}

\begin{figure}
\begin{center}
  \includegraphics{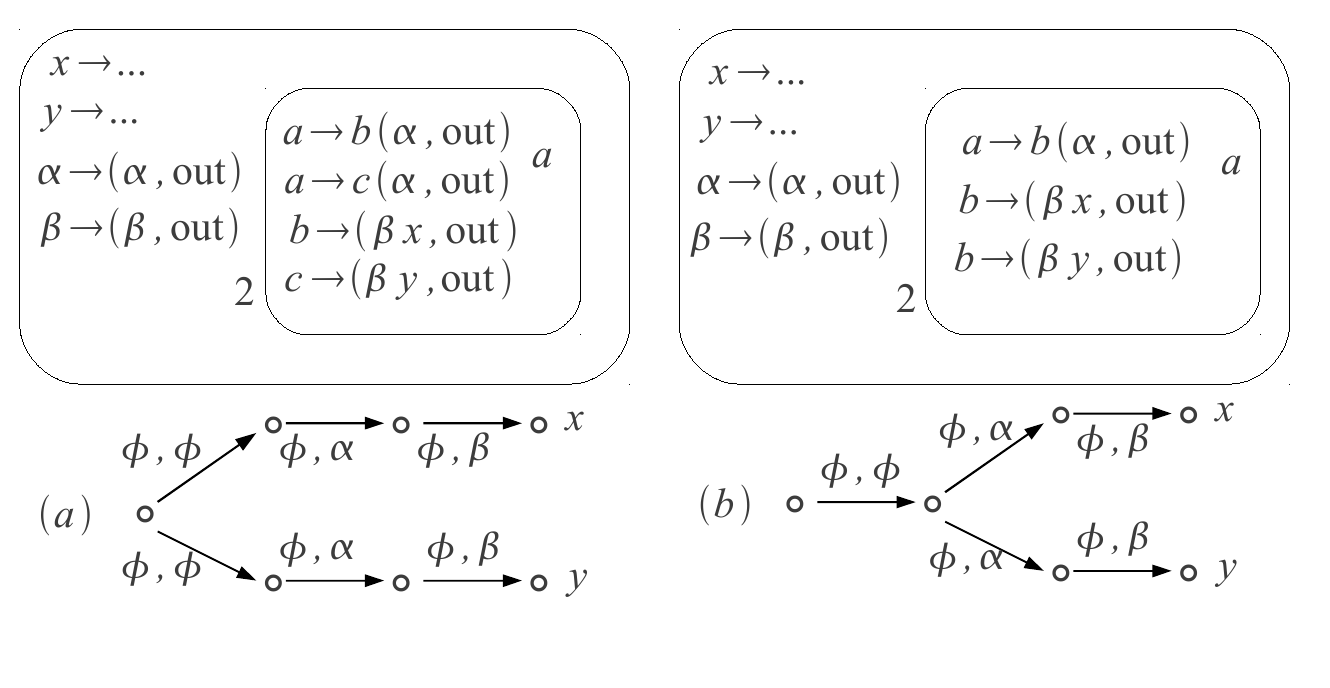}
\end{center}
\caption{Two membrane systems that are test equivalent, but not bisimilar,
together with their graph of transitions.}
\label{fig:testbisim}
\end{figure}

\section{Example of testing scenario}
\label{sec:esempio}
In this section we model a population of individuals that can reproduce both
sexually and asexually. We define different observers to show the expressiveness
of the defined testing framework. It is worth noting that \emph{quantitative}
aspects of systems can be easily expressed. Note that the defined observers are
very specific and are intended to only show the capabilities of the framework
introduced above. In particular, they should not be intended as examples of
verification, because this analysis is to be done by checking the testing
equivalence between the system and its expected behaviour, modelled as a simpler
``specification'' P system.

Most animal species use sexual reproduction to produce offspring, while a
minority of species reproduce asexually by producing clones of the mother. Both
strategies have advantages and weaknesses. During sexual reproduction genes from
two individuals are combined in the offspring that receives genetic material
from both parents involving, in diploid populations, recombination among genes.
Recombination can break up favourable sets of genes accumulated by selection.
Moreover, asexual populations composed by only females can reproduce twice as
fast in each generation than sexual populations, because there is no need to
produce males for ongoing reproduction.

Despite its considerable cost, sexual
reproduction it is still by far the most frequent mode of reproduction in
vertebrates. Asexual reproduction has only been described in less than $0.1\%$
of vertebrate species. In general, it is assumed that sexually reproducing
populations harbour more genetic variation than asexually reproducing
populations, and a high level of genetic variation allows perpetual adaptation
to changing environments.

Particularly, populations in heterogeneous habitats,
threatened by various parasites or under strong competition, have been shown to
have greater genetic variation. For the above reasons a variety of species
(essentially among invertebrates) adopted a mixed strategy which tries to
combine the advantages of both methods \cite{Schon2009}.

In this example we model a simple organism able to reproduce either sexually or
asexually. We consider the individual of the species as diploid with only  a
locus (gene), thus each genotype is composed  by a pair $(a_1, a_2)$ of alleles
which the two chromosomes have for the gene. Moreover, we consider the sex of
individuals, thus each of them is represented by a pair of alleles together with
the symbol, $f$ or $m$, of the sex. The rules controlling the evolution of the
population are {\it  reproduction rules}, either sexual reproduction rules or
asexual, and {\it  death rules}. Each rule has an inhibitor; when the inhibitor
is present the rule cannot be applied.

Consider a set of alleles (values for the single gene) of $k$ elements $\{ v_1,
v_2 , \ldots , v_k\}$. In the following $ a_1, a_2, a_3, a_4$ belong to $\{ v_1,
v_2 , \ldots , v_k\}$. The reproduction rules are the following:

$$
\begin{array}{ll}
1. & a_1 a_2 m \ \ a_3 a_4 f \rightarrow a_1 a_2 m \ \ a_3 a_4 f \ \ a_i a_j s\
|_{\neg no\_sex\_repr}\ \ (i\in\{1,2\}, j\in\{3,4\}, s\in\{m,f\})\\ 
2. & a_1 a_2 f  \rightarrow a_1 a_2 f \ \ a_1 a_2 f\ |_{\neg no\_asex\_repr}\\ 

\end{array}
$$

Note that each rule has its inhibitor ${no\_sex\_repr}$ or ${no\_asex\_repr}$.
The death rules are the following:

$$
\begin{array}{ll}
3. & a_1 a_2 m \rightarrow \lambda\ |_{\neg no\_male\_death}\\

4. & a_1 a_2 f  \rightarrow \lambda\ |_{\neg no\_female\_death}\\ 
\end{array}
$$

We add also rules for females and males which simply survive, without
reproducing or dying:

$$
\begin{array}{ll}
5. & a_1 a_2 m \rightarrow a_1 a_2 m\ |_{\neg no\_male\_life}\\ 
6. & a_1 a_2 f  \rightarrow a_1 a_2 f \ |_{\neg no\_female\_life}\\ 
\end{array}
$$

Finally we use a rule for sending out individuals from the membrane in which the
population evolved. This rule is {\it  promoted} by the promoter $send\_out$:

$$
\begin{array}{ll}
7. & a_1 a_2 m \rightarrow (a_1 a_2 m,\ out)\ |_{send\_out}\\ 
8. & a_1 a_2 f  \rightarrow (a_1 a_2 f,\ out) \ |_{send\_out}\\ 
\end{array}
$$

In the following examples, we consider the membrane system defined above as the
system under test. In each example we define a specific observer that is meant
to test
the evolution forcing certain situations. The observer controls
the system by
sending into it both the initial individuals and the promoters/inhibitors for
constraining the population dynamics.

\begin{example}
\label{esempio1}
Let us consider a population in which the possible alleles for the single locus
are $\{0,1\}$ and an initial population composed by  four males of genotype
$(00)$ and four females of genotype $(01)$. Let us control the population
dynamics by inhibiting both the sexual reproduction and the death of females.
The observer we define, using the \must version, is able to analyse the
following property of the system: ``{\it  After two time units, no female in the
population can differ from the initial ones, and the number of such females is
greater than or equal to the initial female number.}''

Note that naturally the tests express \emph{quantitative} aspects, both on
time and on numbers of individuals. Assume that membrane $1$, i.e.\ the
observer,
initially contains  the element $a$ and that $Inh$ is the set of all inhibitors,
the rules in membrane $1$ are the following:
$$
\begin{array}{lll}
1_0. & a \rightarrow 1 \ ((00m)^4,(01f)^4, no\_sex\_repr, no\_female\_death,\
in_2)\\ 
2_0. & 1 \rightarrow 2 \\
3_0. & 2 \rightarrow 3 \ (Inh \cup\{send\_out\}, \ in_2)\\
4_0. & 3 \rightarrow 4 \\ 
5_0. & a_1 a_2 f \rightarrow  fail \ \ \ \ \ (a_1\not= 0 \vee a_1\not= 1)\\ 
6_0. & 4 \rightarrow 5\\
7_0. & 5 \ (01f)^4 \rightarrow (\omega,\ out) \ |_{\neg fail}\\ 
\end{array}
$$
Rule $1_0$ sends into the membrane under test the initial population (four males
and four females) and the inhibitors for sexual reproduction and death of
females. Rule $2_0$ waits for a time unit, and, after that, Rule $3_0$ sends,
during the second step of the populations evolution, all the inhibitors together
with the promoter $send\_out$.
Rule $4_0$ waits for a time unit to allow the inner membrane to send out all the
individuals. Afterwards, Rule $5_0$ produces a $fail$ if a female different from
the initial ones is present. At the same time Rule $6_0$ increases the counter.
Finally, Rule $7_0$ sends out the $\omega$ symbol only if $fail$ is absent.
\end{example}
\begin{example}\label{esempio2}
Consider the initial population of Example~\ref{esempio1}, under the same
conditions. Again using a \must test, we consider the following property: {\it 
``Given any $k \in \mathbb{N}$, we can check that it is not possible, in $n$
time units, for all $n\in [1,k]$, to produce a female different from the initial
ones.}'' Note that this property is not a for-all statement: we count until the
given $k$ for checking it. This is of course weaker than checking the for-all
statement.

The rules in membrane $1$ are the following:
$$
\begin{array}{lll}
1_0. & a \rightarrow 1 \ block\ ((00m)^4,(01f)^4, no\_sex\_repr,
no\_female\_death,\ in_2)\\ 
2_0. & 1 \rightarrow 2 \\
3_0. & 2 \rightarrow 3 \\ 
&\cdots\\
k_0. & k-1 \rightarrow k \\ 
(k+1)_0. & block \rightarrow block \ |_{\neg k}\\
(k+2)_0. & block \rightarrow 1' \ (Inh \cup\{send\_out\}, \ in_2)\\
1'_0. & 1' \rightarrow 2' \\
2'_0. & a_1 a_2 f \rightarrow  fail \ \ \ \ \ (a_1\not= 0 \vee a_1\not= 1)\\
3'_0. & 2' \rightarrow 3'\\ 
4'_0. & 3' \rightarrow (\omega,\ out) \ |_{\neg fail}\\
\end{array}
$$
Rules from $2_0$ to $k_0$ increase the counter until $k$. At each time unit
either Rule $(k+2)_0$ can be executed, stopping the evolution of the population,
or Rule $(k+1)_0$ can be fired, allowing the population to evolve for one more
step. Rule $(k+1)_0$ is inhibited by $k$, thus when the counter reaches $k$ the
evolution must terminate. Rules from $1'_0$ to $4'_0$ produce a $\omega$ if and
only if, in the final populations there are only females equal to the initial
ones.
\end{example}
\begin{example}\label{esempio3}
Consider the initial population of Example~\ref{esempio1}. This time let
us consider a \may test expressing the following: ``{\it  Without initial
conditions it is possible to have recombination (offspring with different
genotypes with respect to the initial population) after $k$ steps.}''
\par\smallskip
The rules are the following:
$$
\begin{array}{lll}
1_0. & a \rightarrow 1 \ block\ ((00m)^4,(01f)^4, \ in_2)\\
2_0. & 1 \rightarrow 2 \\
3_0. & 2 \rightarrow 3 \\ 
&\cdots\\
k_0. & k-1 \rightarrow k \\ 
(k+1)_0. & k \rightarrow 1' \ (Inh \cup\{send\_out\}, \ in_2)\\
1'_0. & 1' \rightarrow 2' \\
2'_0. & a_1 a_2 f \rightarrow  (\omega,\ out) \ \ \ \ \ (a_1\not= 0 \vee
a_1\not= 1)\\
2'_0. & a_1 a_2 m \rightarrow  (\omega,\ out) \ \ \ \ \ (a_1\not= 0 \vee
a_1\not= 1)\\ 
\end{array}
$$
\end{example}
\begin{example}\label{esempio4}
Consider again the initial population of Example~\ref{esempio1}. Let us define
a \may test expressing: ``{\it  By allowing only the asexual reproduction for $k$
steps,
and then allowing only the sexual
reproduction for the following $k$ steps, it is possible to have recombination
in the final population.}''
\par\smallskip
For this example we need the concept of {\it  antidote}. An antidote is a symbol
able to remove the effect of an inhibitor, usually for an inhibitor $x$, the
antidote is denoted by $anti\_x$. The effect of an antidote is described by
particular rules, the {\it  antidote rules}, which have the form $anti\_x \ x
\rightarrow \lambda$. In this example we assume that, in the membrane under
test, there are the antidote rules for the inhibitors $no\_sex\_repr$ and
$no\_sex\_repr$. The rules are the following:
$$
\begin{array}{lll}
1_0. & a \rightarrow 1 \ block\ ((00m)^4,(01f)^4, no\_sex\_repr,\ in_2)\\
2_0. & 1 \rightarrow 2 \\
&\cdots\\
k_0. & k-1 \rightarrow k \ (anti\_no\_sex\_repr, no\_asex\_repr, \ in_2)\\ 
(k+1)_0. & k \rightarrow k+1\\
(2k)_0. & 2k-1\rightarrow 1' \ (Inh \cup\{send\_out\},\ in_2)\\
1'_0. & 1' \rightarrow 2' \\
2'_0. & a_1 a_2 f \rightarrow  (\omega,\ out) \ \ \ \ \ (a_1\not= 0 \vee
a_1\not= 1)\\
2'_0. & a_1 a_2 m \rightarrow  (\omega,\ out) \ \ \ \ \ (a_1\not= 0 \vee
a_1\not= 1)\\
\end{array}
$$
\end{example}


\section{Conclusions}
The testing machinery defined in \cite{DeNicola1984} and the P Algebra proposed
in \cite{Barbuti2010} inspired us a suitable Process Algebra-based testing
environment for P Systems. On the one hand, the new testing environment shares
with the original one the concepts of observer, running test, successful and
unsuccessful computation, testing preorders/equivalences, allowing us to define
qualitative system properties. On the other hand, differently from the original
one, it results to be suitable also to express {\it  quantitative}
aspects. Such a feature puts in evidence an expected high expressive power
of the framework itself, which needs to be formally studied.

The natural continuation of this work is to find finite decidable
characterizations of testing equivalence of finite state P Algebra terms in
order to perform verification by comparing a system with its expected
behaviour.
Moreover, we plan to extend the testing environment also studying a suitable
version of {\it  fair} testing semantics for P Systems, as well as rephrasing the
testing environment for Spatial P Systems \cite{sps}, with the aim of expressing
quantitative properties involving spatial information, being crucial in the
biological (and not only) domain.

On the biological side, we intend to show as future work the potentials of the
testing framework of having a practical impact, for instance on planning both
in-silico and wet-lab experiments.

\bibliographystyle{abbrv} 
\bibliography{testingpsystems}

\begin{thebibliography}{10}

\bibitem{aguado02}
J.~Aguado, T.~Balanescu, T.~Cowling, M.~Gheorghe, M.~Holcombe, and F.~Ipate.
\newblock {P Systems with Replicated Rewriting and Stream X-machines (Eilenberg
  Machines)}.
\newblock {\em Fundam. Inf.}, 49:17--33, January 2002.

\bibitem{Ahmad2006}
J.~Ahmad, G.~Bernot, J.~Comet, D.~Lime, and O.~Roux.
\newblock {Hybrid Modelling and Dynamical Analysis of Gene Regulatory Networks
  with Delays}.
\newblock {\em ComPlexUs}, 3(4):231--251, 2006.

\bibitem{andrei07}
O.~Andrei, G.~Ciobanu, and D.~Lucanu.
\newblock {A Rewriting Logic Framework for Operational Semantics of Membrane
  Systems}.
\newblock {\em Theor. Comput. Sci.}, 373(3):163--181, 2007.

\bibitem{Barbuti2010}
R.~Barbuti, A.~Maggiolo-Schettini, P.~Milazzo, and D.~Gruska.
\newblock {A Notion of Biological Diagnosability Inspired by the Notion of
  Opacity in Systems Security}.
\newblock {\em Fundamenta Informaticae}, 102(1):19--34, 2010.

\bibitem{sps}
R.~Barbuti, A.~Maggiolo-Schettini, P.~Milazzo, G.~Pardini, and L.~Tesei.
\newblock {Spatial P Systems}.
\newblock {\em Natural Computing}, 2010.
\newblock Received: 26 October 2009 Accepted: 24 February 2010 Published
  online: 24 March 2010.

\bibitem{Barbuti2008a}
R.~Barbuti, A.~Maggiolo-Schettini, P.~Milazzo, and S.~Tini.
\newblock {A P Systems Flat Form Preserving Step-by-step Behaviour}.
\newblock {\em Fundamenta Informaticae}, 87(1):1--34, 2008.

\bibitem{Barbuti2008}
R.~Barbuti, A.~Maggiolo-Schettini, P.~Milazzo, and S.~Tini.
\newblock {Compositional Semantics and Behavioral Equivalences for P Systems}.
\newblock {\em Theoretical Computer Science}, 395(1):77--100, 2008.

\bibitem{Bianco2006}
L.~Bianco and V.~Manca.
\newblock {Encoding-Decoding Transitional Systems for Classes of P systems}.
\newblock In {\em {Proc. of WMC 2005, 6th International Workshop on Membrane
  Computing}}, number 3850 in LNCS, pages 134--143. Springer-Verlag, 2006.

\bibitem{Boreale1999}
M.~Boreale and R.~Pugliese.
\newblock {Basic Observables for Processes}.
\newblock {\em Information and Computation}, 149(1):77--98, 1999.

\bibitem{Bottoni2002}
P.~Bottoni, C.~Mart{\'\i}n-Vide, G.~P{\u{a}}un, and G.~Rozenberg.
\newblock {Membrane Systems with Promoters/Inhibitors}.
\newblock {\em Acta Informatica}, 38(10):695--720, 2002.

\bibitem{ciobanu2006}
G.~Ciobanu, M.~J. P{\'e}rez-Jim{\'e}nez, and G.~Paun, editors.
\newblock {\em {Applications of Membrane Computing}}.
\newblock Natural Computing Series. Springer, 2006.

\bibitem{dang06}
Z.~Dang, C.~Li, O.~H. Ibarra, and G.~Xie.
\newblock {On the Decidability of Model-checking for P Systems}.
\newblock {\em J. Autom. Lang. Comb.}, 11:279--298, January 2006.

\bibitem{DeNicola1984}
R.~De~Nicola and M.~Hennessy.
\newblock {Testing Equivalences for Processes}.
\newblock {\em Theoretical Computer Science}, 34(1-2):83--133, 1984.

\bibitem{ghe09}
M.~Gheorghe and F.~Ipate.
\newblock {On Testing P Systems}.
\newblock In D.~Corne, P.~Frisco, G.~Paun, G.~Rozenberg, and A.~Salomaa,
  editors, {\em {Membrane Computing}}, volume 5391 of {\em Lecture Notes in
  Computer Science}, pages 204--216. 2009.

\bibitem{ghe10}
M.~Gheorghe and F.~Ipate.
\newblock {Testing Based on P Systems - An Overview}.
\newblock In {\em {Proceedings of the 11th International Conference on Membrane
  Computing}}, CMC'10, pages 3--6, Berlin, Heidelberg, 2010. Springer-Verlag.

\bibitem{ipate09}
F.~Ipate and M.~Gheorghe.
\newblock {Testing Non-deterministic Stream X-machine Models and P systems}.
\newblock {\em Electron. Notes Theor. Comput. Sci.}, 227:113--126, January
  2009.

\bibitem{ipate10}
F.~Ipate, M.~Gheorghe, and R.~Lefticaru.
\newblock {Test Generation From P systems Using Model Checking}.
\newblock {\em Journal of Logic and Algebraic Programming}, 79(6):350--362,
  2010.
\newblock Membrane computing and programming.

\bibitem{Ipate2011}
F.~Ipate, R.~Lefticaru, and C.~Tudose.
\newblock {Formal Verification of P Systems using SPIN}.
\newblock {\em International Journal of Foundations of Computer Science},
  22(1):133--142, 2011.

\bibitem{NC95}
V.~Natarajan and R.~Cleaveland.
\newblock {Divergence and Fair Testing}.
\newblock In {\em Proceedings of the 22nd International Colloquium on Automata,
  Languages and Programming}, ICALP '95, pages 648--659, 1995.

\bibitem{Paun2002a}
G.~P{\v{a}}un.
\newblock {\em {Membrane Computing. An Introduction}}.
\newblock Springer-Verlag, 2002.

\bibitem{Plotkin2004}
G.~D. Plotkin.
\newblock {A Structural Approach to Operational Semantics}.
\newblock {\em J. Log. Algebr. Program.}, 60-61:17--139, 2004.

\bibitem{Paun2002}
G.~P\v{a}un and G.~Rozenberg.
\newblock {A Guide to Membrane Computing}.
\newblock {\em Theoretical Computer Science}, 287(1):73--100, 2002.

\bibitem{RV07}
A.~Rensink and W.~Vogler.
\newblock Fair testing.

\bibitem{Schon2009}
I.~Sch{\"o}n, K.~Martens, and P.~van Dijk.
\newblock {\em {Lost Sex: The Evolutionary Biology of Parthenogenesis}}.
\newblock Springer Verlag, 2009.

\end{thebibliography}
\end{document}